\documentclass[11pt]{article}             
\usepackage{osid}                         %

\title{Structure constants of $su(2S+1)$ algebra and the
 decomplexification of the Liouville-von Neumann equation}
\author{E. A. Ivanchenko \\{\footnotesize\it Kharkov Institute of Physics and Technology,
 Institute for Theoretical Physics, 1 Akademicheskaya Str., 61108 Kharkov, Ukraine \\
E-mail: yevgeny@kipt.kharkov.ua}}

\renewcommand{\RHD}{E. A. Ivanchenko. Structure constants of $su(2S+1)$ algebra
and ...}

\begin{document}

\maketitle
\begin{abstract}
     The analytic formulas for structure constants of $su(2S+1)$ algebra in
terms of $3jm$ and $6j$ symbols of $su(2)$ have been derived for the
decomplexification of the Liouville-von Neumann equation.
\end{abstract}

\section{Introduction}
Let $ \{ C_1^S,C_2^S,...,C_n^S\}$  be a base of su(2S+1) algebra, where
$S=1/2,1,3/2,... $ is the spin quantum number, $n=(2S+1)^2-1$. We have according to
\cite{Ivanchenko:kimura&kossakowski}
\begin{equation}\label{equation1}
C_i^SC_j^S=\frac{c}{d}E\delta_{ij}+z_{ijk}^SC_k^S,~{\rm
Tr\,}C_i^S=0,~ {\rm Tr \,}C_i^SC_j^S=c \delta_{ij},
\end{equation}
\begin{equation}\label{equation2}
z_{ijk}^S=g_{ijk}^S+ie_{ijk}^S,
\end{equation}
hence
\begin{equation}\label{equation3}
-i[C_i^S,C_j^S]=2e_{ijk}^SC_k^S,
~\{C_i^S,C_j^S\}=\frac{c}{d}E\delta_{ij}+2g_{ijk}^SC_k^S,
\end{equation}
\begin{equation}\label{equation4}
e_{ijk}^S=\frac{1}{2ic}{\rm Tr\,}[C_i^S,C_j^S]C_k^S,
\end{equation}
\begin{equation}\label{equation5}
g_{ijk}^S=\frac{1}{2c}{\rm Tr\,}\{C_i^S,C_j^S\}C_k^S,
\end{equation} where $d=2S+1$, $E$ is the unit
matrix in  dimension $d \times d$, $c$ is a constant, {\rm Tr\,} is a symbol for
trace. It is easy to see that the structure constants $e_{ijk}^S$ and $g_{ijk}^S$ are
completely antisymmetric  and symmetric in the displacement of any pair of indices.

\section{Hermitian basis}
In order to calculate the structure constants we have to choose the basis.  The basis
is based on linear combinations of irreducible tensor operators. The matrix
representations of  irreducible tensor operators $T_{k,q}^S$
\cite{Ivanchenko:varsalolovic&moskalev&chersonskij} can be calculated  using the
Wigner $3jm$ symbols:
\begin{equation}\label{equation6}
T_{k,q}^S=\sqrt{(2S+1)(2k+1)}\sum^S_{m,m'=-S}(-1)^{S-m} \left(
^{\,\,S\,\,\,\,\,\,k\,\,\,S}_{-m\,\,q\,\,\,m'} \right) |S,m><S,m'|,
\end{equation}
where $0\leq k \leq 2S$, and $-k\leq q \leq k$ in steps of 1. The normalization is
such that $T_{0,0}^S=E$. It is known that the Cartesian product operators $S_x,S_y$,
and $S_z$ for  spin $S=\frac{1}{2}$ are Hermitian and can be calculated from
irreducible tensor operators  \cite{ernst&bodenhausen&wokaun}
\begin{equation}\label{equation7}
S_x^{\frac{1}{2}}=\frac{1}{2\sqrt{2}}(T_{1,-1}^{\frac{1}{2}}-T_{1,1}^{\frac{1}{2}})=
\frac{1}{2}\left(\!\!\!%
\begin{array}{cc}
  0 & 1 \\
  1 & 0 \\
\end{array}%
\!\!\!\right),
\end{equation}
\begin{equation}\label{equation8}
S_y^{\frac{1}{2}}=\frac{i}{2\sqrt{2}}(T_{1,-1}^{\frac{1}{2}}+T_{1,1}^{\frac{1}{2}})=
\frac{1}{2}\left(\!\!\!%
\begin{array}{cc}
  0 & -i \\
  i & 0 \\
\end{array}%
\!\!\!\right),
\end{equation}
\begin{equation}\label{equation9}
S_z^{\frac{1}{2}}=\frac{1}{2}T_{1,0}^{\frac{1}{2}}=\frac{1}{2}\left(\!\!\!%
\begin{array}{cc}
  1 & 0 \\
  0 & -1 \\
\end{array}%
\!\!\!\right).
\end{equation}
Allard and H\"{a}rd \cite{Ivanchenko:allard&hard} have formed linear combinations of
the irreducible tensor operators not only for single-quantum coherences, but also for
all coherences according to
\begin{equation}\label{equation10}
 C_{k,qx}^S =\sqrt{\frac{S(S+1)}{6}}(T_{k,-q}^S+(-1)^qT_{k,q}^S),~q\neq0,
\end{equation}
\begin{equation}\label{equation11}
C_{k,qy}^S=i\sqrt{\frac{S(S+1)}{6}}(T_{k,-q}^S-(-1)^qT_{k,q}^S),~q\neq0,
\end{equation}
\begin{equation}\label{equation12}
C_{k,z}^S=\sqrt{\frac{S(S+1)}{3}}T_{k,0}^S,~q=0,k\geq1,
\end{equation}
\begin{equation}\label{equation13}
C_{0,z}^S=\sqrt{\frac{S(S+1)}{3}}E,
\end{equation}
where $1\leq k \leq 2S$, and $1\leq q \leq k$  in $C_{k,qx}^S,C_{k,qy}^S$ and $1\leq
k \leq 2S$,
 in $C_{k,z}^S$ in steps of 1. The matrices~(\ref{equation10},\ref{equation11},\ref{equation12}) are traceless and their number is equal to
  $(2S+1)^2-1$.  Using
 the well-known relations for the irreducible tensor operators
\begin{equation}\label{equation14}
(T_{k,q}^S)^+=(-1)^qT_{k,-q},
\end{equation}
 we can see that  matrices~~(\ref{equation10},\ref{equation11},\ref{equation12}) are Hermitian.
Using the formula from  \cite{Ivanchenko:varsalolovic&moskalev&chersonskij}
\begin{equation}\label{equation15}
 {\rm Tr\,} T_{k,q}^ST_{k',q'}^S=(-1)^q(2S+1)\delta_{k,k'}\delta_{q,-q'}
\end{equation}
it is easy to show that the basis is normalized so that $S_x=C_{1,x}^S$,
$S_y=C_{1,y}^S$, $S_z=C_{1,z}^S$,  irrespective of the spin quantum number $S$, i.e.
\begin{equation}\label{equation16}
(C_r,C_s)={\rm Tr\,}C_rC_s=\delta_{r,s}\frac{S(S+1)(2S+1)}{3}.
\end{equation}
The set~(\ref{equation10}-\ref{equation13}) is complete. The matrices $C_{k,z}$ are
diagonal
\begin{equation}\label{equation17}
[C_{k,z}^S,C_{k',z}^S]=0.
\end{equation}
There also exist the other useful bases \cite{Ivanchenko:bertmann&krammer}. From the
physical point of view, for some applications the basis \cite{Ivanchenko:allard&hard}
is preferred.
\section{Analytic formulas for structure constants}
There  are 27 combinations  in threes including the repetitions: $XX'X''$, $XX'Y''$,
$XX'Z''$\dots, where $X=C_{k,qx}^S$,  $X'=C_{k',q'x}^S$, $Y''=C_{k'',q''y}^S$,
$Z''=C_{k'',z}^S$ and so on. The use of the symmetrical properties of the Wigner
$3jm$ symbols and the formula $\textbf{2}.\textbf{4}.(23)$
\cite{Ivanchenko:varsalolovic&moskalev&chersonskij}
\begin{eqnarray}\label{equation18}
{\rm Tr\,}T_{k,q}^S T_{k',q'}^ST_{k'',q''}^S=
(-1)^{2S+k+k'+k''}(2S+1)^{\frac{3}{2}}\nonumber\\
\qquad {}[(2k+1)(2k'+1)(2k''+1)]^{\frac{1}{2}} \{
^{k\,\,\,k'\,\,\,k''}_{S\;\,S\;\;S}\} \left(^{k\;\,k'\;k''}_{q\;\,q'\;q''} \right),
\end{eqnarray}
where $\{ ^{k\,\,\,k'\,\,\,k''}_{S\;\,S\;\;S} \}$ is the $6j$ symbol, allows us after
substitution of ~(\ref{equation10},\ref{equation11},\ref{equation12})
in~(\ref{equation4},\ref{equation5}) , to calculate all structure constants of
$su(2S+1)$ algebra. Let us introduce the function
\begin{equation}\label{equation19}
F(k,k',k'',S)=\frac{(-1)^{2S}}{\sqrt3}\sqrt{S(S+1)(2S+1) (2k+1)(2k'+1)(2k''+1)}\,\{
^{k\,\,\,k'\,\,\,k''}_{S\;\,S\;\;S} \}.
\end{equation}
All antisymmetric structure constants are zero for $K=k+k'+k''$ even and nonvanishing
 antisymmetric structure constants in terms of $3jm$ and $6j$ symbols have the
explicit form  are presented by
formulas~(\ref{equation20},\ref{equation21},\ref{equation22}) for $K$ odd:
\begin{equation}\label{equation20}
e_{XX'Y''}^S = -\frac{F}{\sqrt{2}}\!\!\left[(-\!1)^q
\left(^{k\,\,\,k'\,\,\,k''}_{q-q'-q''}\right)+
(-\!1)^{q'}\left(^{k\;\;\;\,k'\;\;\,k''}_{\!-q\,\,\,q'-q''}\right)+(-\!1)^{q''}
\left(^{k\,\,\,k'\,\,\,k''}_{q\;\;q'-q''}\right) \right],
\end{equation}

\begin{equation}\label{equation21}
e_{YY'Y''}^S =
\frac{F}{\sqrt{2}}\!\!\left[(-\!1)^q\left(^{k\;\;\;\,k'\,\,\,k''}_{\!-q\;\,q'\;q''}
\right)+(-\!1)^{q'}\left(^{k\,\,\,\,k'\,\,\,k''}_{q\,-q'\;q''}\right)+
(-\!1)^{q''}\left(^{k\,\,\,\,k'\,\,\,k''}_{q\;\;q'-q''} \right)\right],
\end{equation}
\begin{equation}\label{equation22}
 e_{XY'Z''}^S =
 -F(-\!1)^q\left(^{k\,\,\,\,k'\,\,\,k''}_{q\,\,-q'\,\,0}\right).
\end{equation}

All symmetric structure constants are zero for $K=k+k'+k''$ odd and nonvanishing
 symmetric structure constants in terms of $3jm$ and $6j$ symbols have the
explicit form  are presented by
formulas~(\ref{equation23},\ref{equation24},\ref{equation25}) for $K$ even:
\begin{equation}\label{equation23}
g_{XX'X''}^S =
\frac{F}{\sqrt{2}}\!\!\left[(-\!1)^q\left(^{k\;\;\;\,k'\,\,\,k''}_{q\,\,-q'\,-q''}
\right)+(-\!1)^{q'}\left(^{k\;\;\;\,k'\,\,\,k''}_{\!-q\,\,q'\,-q''}\right)+
(-\!1)^{q''}\left(^{k\,\,\,\,k'\,\,\,k''}_{q\;\;q'-q''} \right)\right],
\end{equation}

\begin{equation}\label{equation24}
g_{XY'Y''}^S =
\frac{F}{\sqrt{2}}\!\!\left[-(-\!1)^q\left(^{k\,\,\,\,k'\;\;\,k''}_{q\,\,-q'\,-q''}
\right)+(-\!1)^{q'}\left(^{k\;\,\;k'\;\;\,k''}_{\!-q\,\,q'\,-q''}\right)+
(-\!1)^{q''}\left(^{k\,\,\,\,\,\,k'\,\,\,\,k''}_{\!-q\,-q'\;q''} \right)\right],
\end{equation}

\begin{equation}\label{equation25}
 g_{XX'Z''}^S =g_{YY'Z''}^S= F(-\!1)^q
 \left(^{k\;\;\;k'\,\,\,k''}_{q\,\,-q'\;0}\right),~
 g_{ZZ'Z''}^S = F(-\!1)^q\left(^{k\;\,k'\,k''}_{0\,\,\,0\,\,\,0} \right).
\end{equation}
We have in $X,Y$  \,$1\leq k,k',k'' \leq 2S$,  $1\leq q \leq k, 1\leq q' \leq k',
1\leq q'' \leq k''$ and in $Z$  \, $1\leq k,k',k'' \leq 2S$ in steps of 1. \\
The direct calculation confirms that the structure constants $e_{ijk}^S$ and
$g_{ijk}^S$ are completely antisymmetric  and symmetric in the displacement of any
pair of operators. In other words it is $e_{XX'Y''}^S=-e_{XY''X'}^S$, $g_{XX'Z''}^S
=g_{YY'Z''}^S$ and so on.

\section{Decomplexification of the Liouville-von Neumann equation}
The  structure constants of $su(2S+1)$ algebra have wide physical applications.
 Let us consider the Liouville-von Neumann
equation for the density matrix  $ \rho $, describing the dynamics of a  system. It
has the form
\begin{equation}\label{equation26}
  i\partial_t\rho=[\hat{H},\rho],~ \rho(t=0)=\rho_0,~ \rho^+=\rho,~{\rm Tr\,}\rho=1,
\end{equation}
 where
  $\hat{H}$ is the Hamiltonian of the  system.\\
  \indent The state of a $d$-dimensional quantum system (qudit) is usually described by the  $d \times d$
  density matrix $\rho$.
 Let us present the solution of the equation~ (\ref{equation26}) for one qudit as
$\rho=\frac{1}{(2S_1+1)\sqrt{\frac{S_1(S_1+1)}{3}}}R_{\alpha}C_{\alpha}^{S_1}$, $
R_0=1$,  $\hat{H}=\frac{1}{2} h_{\beta}C_{\beta}^{S_1}$. We multiply~
(\ref{equation26}) by $C_{\gamma}^{S_1}$ and execute the operation ${\rm Tr\,}$,
where by definition $ {\rm Tr\,} \rho C_{\alpha}^{S_1} =
\sqrt{\frac{S_1(S_1+1)}{3}}R_{\alpha},\, C_{\alpha}^{S_1} \in
\{C_{k,qx}^{S_1},C_{k,qy}^{S_1},C_{k,z}^{S_1},C_{0,z}^{S_1} \} $. Hereinafter, the
summation is made over the repeating Greek indices on the
set~(\ref{equation10}-\ref{equation13}), and over Latin indices on the
set~(\ref{equation10},\ref{equation11},\ref{equation12}) and then on $k$, and then on
$q$. The Liouville-von Neumann equation takes on the real form in terms of the
functions $R _ {j} $ as  a closed system of differential equations for the set of
initial conditions \cite{Ivanchenko:hioe&eberly}:
\begin{equation}\label{equation27}
\partial_tR_{l}=e_{ijl}^{S_1} h_iR_{j}.
\end{equation}
 The length of the generalized Bloch vector $ b^{S_1} $   is conserved under
unitary evolution
\begin{equation}\label{equation28}
b^{S_1} =\sqrt{R^2_{m}}.
\end{equation}
 \indent For two different
coupled qudits we have $\hat{H}=\frac{1}{2} h_{\alpha \beta}C_{\alpha}^{S_1} \otimes
C_{\beta}^{S_2}$,
\begin{equation}\label{equation29}
  \rho=\frac{3}{(2S_1+1)(2S_2+1)
  \sqrt{S_1(S_1+1)S_2(S_2+1)}}R_{\gamma\delta}C_{\gamma}^{S_1}\otimes C_{\delta}^{S_2},
 \end{equation}
 $R_{00}=1$,
where $ \otimes $ is the symbol of direct product.\\
  The dynamic equation~ (\ref{equation26}) for two different qudits takes on the real
form in terms of the functions $R _ {m0},R _ {0m},R _ {mn} $ as a closed system of
differential equations~(\ref{equation30},\ref{equation31},\ref{equation32}) for the
set of initial conditions:
\begin{equation}\label{equation30}
  \partial_tR_{m0}=\sqrt{\frac{S_2(S_2+1)}{3}}e_{pim}^{S_1}(h_{p0}R_{i0}+h_{pl}R_{il}),
    \end{equation}
\begin{equation}\label{equation31}
  \partial_tR_{0m}=\sqrt{\frac{S_1(S_1+1)}{3}}e_{pim}^{S_2}(h_{0p}R_{0i}+h_{lp}R_{li}),
  \end{equation}
  \begin{eqnarray}\label{equation32}
  \partial_tR_{mn} &= e_{pim}^{S_1}\left[\sqrt{\frac{S_2(S_2+1)}{3}}(h_{pn}R_{i0}+h_{p0}R_{in})+
  g_{rln}^{S_2}h_{pr}R_{il}\right]+ \nonumber\\
&
 e_{pin}^{S_2}\left[\sqrt{\frac{S_1(S_1+1)}{3}}(h_{mp}R_{0i}+
  h_{0p}R_{mi})+
  g_{rlm}^{S_1}h_{rp}R_{li}\right],
  \end{eqnarray}
  where by definition\\
\begin{equation}\label{equation33}
{\rm Tr\,}\rho C_{\alpha}^{S_1} \otimes
C_{\beta}^{S_2}=\frac{1}{3}\sqrt{S_1(S_1+1)S_2(S_2+1)}R_{\alpha\beta},
\end{equation}
and  $ C_{\beta}^{S_2} \in
\{C_{k,qx}^{S_2},C_{k,qy}^{S_2},C_{k,z}^{S_2},C_{0,z}^{S_2}\} $ .
 The  functions $R_{m0}, R_{0m}$  describe the individual qudits
and the  functions $R_{mn}$ define their correlations.

 The length of the generalized Bloch vector $ b^{S_1S_2} $   is conserved under unitary
evolution
\begin{equation}\label{equation34}
b^{S_1S_2} =\sqrt{R^2_{m0}+R^2_{0m}+R^2_{mn}}.
\end{equation}
 The set of equations for 3 qubits has been obtained in
\cite{Ivanchenko:ivanchenko}. The  dynamic equations for 3 different qudits will be
presented elsewhere.

\section{Conclusion}

It is not necessary for the basis to be Hermitian since the results of calculations
are independent of the choice of base, but there is the main advantage with the
Hermitian basis. It is that the Liouville-von Neuman equation not involve  any
complex numbers and can be solved using real algebra. This is not true for
non-Hermitian bases. Real algebra makes numerical calculations faster and simplifies
the interpretation  of the
equation system.\\
This basis forms a natural basis for calculations on coupled spin systems
(multipartite systems) \cite{Ivanchenko:allard&hard} because all the single-spin
operators are part of the complete basis when the unit operator is  part of the
single-spin basis.\\
The  convolutions of  structure constants \cite{Ivanchenko:macfarlane&sudbery&weisz}
can give rise to the additional relations
between  $3jm$ and $6j$ symbols.\\
The relationship between the basis \cite{Ivanchenko:allard&hard} and the Gell-Mann
basis for spin 1 has been presented in the appendix.

\section*{Appendix}

 The matrix representation of the complete set of the Hermitian  operators for spin 1
 (for qutrit)
 has the form

 $ C_{0,z}^1=C_0=\sqrt{2/3}\left( \begin{array}{ccc}
  1 & 0 & 0 \\
  0 & 1 & 0 \\
  0 & 0 & 1 \\
\end{array}\right),~ C_{1,x}^1=C_1= 1/\sqrt{2}\left( \begin{array}{ccc}
  0 & 1 & 0 \\
  1 & 0 & 1 \\
  0 & 1 & 0 \\
\end{array}\right),\\
 C_{1,y}^1=C_2=i/\sqrt{2}\left( \begin{array}{ccc}
  0 & -1 & 0 \\
  1 & 0 & -1 \\
  0 & 1 & 0 \\
\end{array}\right),~
 C_{1,z}^1=C_3= \left( \begin{array}{ccc}
  1 & 0 & 0 \\
  0 & 0 & 0 \\
  0 & 0 & -1 \\
\end{array}\right),\\
 C_{2,2y}^1=C_4=i\left( \begin{array}{ccc}
  0 & 0 & -1 \\
  0 & 0 & 0 \\
  1 & 0 & 0 \\
\end{array}\right),~ C_{1,2y}^1=C_5= i/\sqrt{2}\left( \begin{array}{ccc}
  0 & -1 & 0 \\
  1 & 0 & 1 \\
  0 & -1 & 0 \\
\end{array}\right),\\
 C_{2,z}^1=C_6=1/\sqrt{3}\left( \begin{array}{ccc}
  1 & 0 & 0 \\
  0 & -2 & 0 \\
  0 & 0 & 1 \\
\end{array}\right),~ C_{2,x}^1=C_7=1/\sqrt{2} \left( \begin{array}{ccc}
  0 & 1 & 0 \\
  1 & 0 & -1 \\
  0 & -1 & 0 \\
\end{array}\right),\\
 C_{2,2x}^1=C_8= \left( \begin{array}{ccc}
  0 & 0 & 1 \\
  0 & 0 & 0 \\
  1 & 0 & 0 \\
\end{array}\right) $.\\

 These matrices are traceless ${\rm Tr\,}C_a=0$ and orthogonal
${\rm Tr\,} C_a C_b =2 \delta_ {ab} $, $ 1 \leq a, b\leq 8$. The relationship between
the basis $C_a$ \cite{Ivanchenko:allard&hard} and the
Gell-Mann basis  ${\lambda_a}$ is   the following:\\
$C_1=1/\sqrt{2}(\lambda_1+\lambda_6), C_2=1/\sqrt{2}(\lambda_2+\lambda_7),
C_3=1/2\lambda_3+\sqrt{3}/2\lambda_8, C_4=\lambda_5,
C_5=1/\sqrt{2}(\lambda_2-\lambda_7), C_6=\sqrt{3}/2\lambda_3-1/2\lambda_8,
C_7=1/\sqrt{2}(\lambda_1-\lambda_6), C_8=\lambda_4.$\\
The antisymmetric (symmetric) structure constants $e_{abc}$ ($g_{abc}$) are
correspondingly equal to: $e_{123}=e_{158}=e_{254}=e_{278}=e_{375}=e_{471}=1/2,\
e_{156}=e_{672}=\sqrt{3}/2,\ e_{348}=-1$;
$g_{336}=g_{446}=-g_{666}=g_{688}=1/\sqrt{3},\ g_{556}=g_{116}=g_{226}=g_{677}=-1/{2
\sqrt{3}},\ g_{235}=g_{118}=g_{558}=g_{124}=g_{137}=-g_{228}=-g_{778}=-g_{475}=1/2$.
Hence, we have $C_aC_b=2/3\delta_{ab}+g_{abc}C_c+ie_{abc}C_c$.


The author is grateful to A. A. Zippa  for fruitful discussions
 and constant invaluable support.


\begin{thebibliography}{99}

\footnotesize
\bibitem{Ivanchenko:kimura&kossakowski} G.~Kimura and A.~Kossakowski,
  Open Systems \& Information Dynamics. {\bf 12}, 207 (2005);  quant-ph/0408014.

\bibitem{Ivanchenko:varsalolovic&moskalev&chersonskij} D. A. Varshalovich  A. N. Moskalev,
 and V. K. Khersonskii,  {\em Quantum Theory of Angular
Momentum\/} (Leningrad: "Nauka" edition) 1975.


\bibitem{ernst&bodenhausen&wokaun}
R.R. Ernst, G. Bodenhausen, and  A. Wokaun, {\em Principles of Nuclear Magnetic
Resonance
 in One and Two Dimensions\/},
  Oxford Univ. Press,
Oxford, 1987.

\bibitem{Ivanchenko:allard&hard}
 P. Allard and T. H\"{a}rd,  J.~Mag. Resonance, {\bf 153}, 15, (2001).

\bibitem{Ivanchenko:bertmann&krammer} R. A. Bertmann and P. Krammer,
   ArXiv:0706.1743 (2007).

\bibitem{Ivanchenko:hioe&eberly}
 F. T. Hioe and J. H. Eberly,  Phys. Rev. Letters, {\bf 47}, 838, (1981).

\bibitem{Ivanchenko:ivanchenko} E. A. Ivanchenko, Low Temp. Physics, {\bf 33}(4), 336,
(2007);
 quant-ph/0610176.

 \bibitem{Ivanchenko:macfarlane&sudbery&weisz}
 A. J. Macfarlane, A. Sudbery and P. H. Weisz, Commun. math. Phys.,
  {\bf 11},  77 (1968).
\end{thebibliography}
\end{document}